\documentstyle[epsf]{article}
\newcommand{\ee}{\end{equation}}
\newcommand{\be}{\begin{equation}}
\begin{document}
\begin{center}
{\bf
Two-time-scale relaxation towards
thermal equilibrium of the enigmatic  piston }
\vskip 3mm
{\rm Christian Gruber and S\'everine Pache}
\vskip 1mm
{\it Institut de Physique Th\'{e}orique, Ecole Polytechnique
F\'ed\'erale de Lausanne, CH-1015 Lausanne, Switzerland}
\vskip 3mm
{\rm Annick Lesne} 
\vskip 1mm
{\it Laboratoire de Physique Th\'eorique 
des Liquides, 
Universit\'e Pierre et Marie Curie,  
Case  121,\\ 4 Place Jussieu, 75252
 Paris Cedex 05,
France\vskip 3mm}
\date{\today}
\end{center}

\vskip 10mm

\begin{abstract}
{\small
We investigate the evolution of a system composed of 
$N$ non-interacting point 
particles of mass $m$ in a container divided into 
two chambers by a movable adiabatic piston
of mass $M\gg m$.
Using a two-time-scale perturbation approach
in terms of the small parameter $\alpha=2m/(M+m)$,
we show that the evolution towards thermal equilibrium
proceeds in two stages.
The  first stage
is a fast, deterministic, adiabatic
relaxation towards mechanical equilibrium.
The second stage, which takes place at times ${\cal O}(M)$,
is a slow fluctuation-driven, diathermic relaxation
towards thermal equilibrium.
A very simple equation is derived which shows that 
in the second stage, the position of the piston
is given by $X_M(t)=L[1/2-\xi(\alpha t)]$
where the function $\xi$ is independent of $M$.
Numerical simulations support the assumptions underlying
our analytical  derivations and illustrate
 the large mass range in which the picture holds.

}
\end{abstract}

\vskip 10mm
\section{Introduction}

The ``adiabatic'' piston problem is a well-known controversial
example of thermodynamics \cite{callen}.
An isolated cylinder contains two identical fluids which initially
are in different equilibrium states and which are 
separated by an adiabatic fixed piston.
The whole system remains therefore in equilibrium.
The problem is then to predict the final state to which the system will 
evolve when the constraint fixing the piston is released.
Although it is a very old problem since it was discussed to measure
experimentally the ratio $c_p/c_v$ of gases already before 1940 \cite{clark}
\cite{lange}, it still remains a controversial
 question since it shows that the two laws of equilibrium thermodynamics
are not sufficient to predict the final state.

\vskip 3mm
Recently,
especially after the work on the second law of
thermodynamics by E. Lieb and J. Ynvagson \cite{lieb1} and the talk of E. Lieb at
the StatPhys 20 meeting in 1998 \cite{lieb2},
 this problem has attracted renewed interest.
It was first realized that it is  a standard example
where one is forced to apply non-equilibrium thermodynamics since the final
state may depend on the values of the friction coefficients \cite{gruber}.
Then a very simple model was considered to investigate the 
evolution from a microscopical point of view \cite{PG}.
This system consists of $N$ non-interacting particles in a cylinder
of length $L$ and cross-section $A$.
It is divided into two compartments containing respectively
$N^-$ and $N^+$ particles  ($N=N^-+N^+$) by an adiabatic
(i.e. no internal 
degrees of freedom) piston of mass $M$.
The dynamics is defined by the condition that the 
piston is constrained to move without friction along the $x$-axis
and the particles make purely elastic collisions
on the boundaries and on the piston. Without loss of generality,
we can assume that all the particles have velocities parallel to
the $x$-axis and thus we are led formally to a one-dimensional problem
 (except for normalization).
Therefore, given that the velocities of a particle and the piston
are $v$ and $V$ before they collide, then after
 the collision the velocities will  be $v^{\prime}$ and $V^{\prime}$,
with:
\be\label{coll}
v^{\prime}=2V-v+\alpha(v-V)
\hskip 20mm V^{\prime}=V+\alpha(v-V)
\hskip 5mm{\rm where}\hskip 5mm
\alpha={2m\over M+m}
\ee
For physical situations where $m\ll M$,
this model was investigated in \cite{GP} using Boltzmann equation
and a perturbation expansion of the velocity distribution 
function of the piston $\Phi_{\epsilon}(V,t)$  in powers
of $\epsilon=\sqrt{m/M}$. 
For an infinite cylinder  $(L=\infty)$ and equal pressures
on both sides of the piston ($p^-=p^+$), it was shown that the stationary
solution of the Boltzmann equation gives a constant  velocity for the piston
$V_{st}=(1/4M)\;
\sqrt{2\pi k_Bm}\;(\sqrt{T^+}-\sqrt{T^-})+{\cal O}(m/M)$ towards
the high temperature domain.
It was thus concluded that stochastic motion together with space asymmetry
(temperature difference) 
implies a macroscopic motion.
Using qualitative arguments and numerical simulations, the case of a finite
cylinder was investigated in \cite{frachebourg}.
It was thus realized that the evolution takes place in two or three stages.
In a first stage, the evolution appears to be deterministic and adiabatic;
this first stage proceeds until mechanical equilibrium
is reached ($p^-=p^+$ but $T^-\neq T^+$).
In the second stage which takes place on much longer 
time scale, the simulations showed that the motion of the piston is stochastic
and proceeds  with exchange of heat through the piston until
thermal equilibrium is reached where $T^-=T^+$,
i.e. the piston which was adiabatic when fixed and during the first stage
becomes heat-conducting under fluctuations.
This explains in particular the results mentioned above for the infinite
cylinder with $p^-=p^+$. In the third stage, under the stochastic
motion of the piston, 
the velocity distribution 
functions of the fluids tend to Maxwellian distributions.

\vskip 3mm
Recently we have studied the adiabatic piston problem in the thermodynamic
limit for the piston, i.e. by considering the limit where $L$ is fixed but 
the area $A$ of the cylinder, the mass $M$ of the piston and the number
$N^{\pm}$  of fluid
particles  tend to infinity with $M/A$ and $R^{\pm}=mN^{\pm}/M$
fixed \cite{GPL}.
Starting from Liouville equation, it was shown 
that in this thermodynamic limit, the motion of the piston is adiabatic
and deterministic, i.e. $\langle V^n\rangle_t=\langle V\rangle_t^n$.
Introducing at this point simplifying assumptions
(see Assumptions 2 and 3 below), we obtained a system of 
autonomous equations from which we concluded that the system 
evolves towards a state of mechanical equilibrium where the pressures
are equal but the temperatures different. Furthermore, 
numerical simulations were conducted which showed that
 the motion depends strongly on $R^{\pm}$
for $R^{\pm}<1$  but tends to be independent of $R^{\pm}$ for $R^{\pm}>10$.
For $R^{\pm}<1$, the evolution is very weakly damped and 
the period of oscillations, which depends on    $R^{\pm}$,
coincide with the period computed with our equations, and with the
period obtained from thermodynamics assuming
adiabatic oscillations \cite{clark} \cite{lange}.
On the other hand for $R^{\pm}>10$, the motion is strongly damped and is 
independent of $R^{\pm}$.
An equation was thus conjectured to describe the evolution for $R^{\pm}$
large enough. It should be noticed that these two types of regimes,
 i.e. weak {\it vs} strong damping, have been observed
 experimentally \cite{lange}, as well as in the previous numerical simulations
 for the simple piston where the evolution is necessarily 
adiabatic \cite{morris}. Let us also stress that in this thermodynamic limit,
 we could deduce from Liouville equation the factorization property
of the joint velocity distribution for the piston and 
a particle at the piston surface.

\vskip 3mm
In the present work we investigate the motion of the piston
with $M$ finite but $M\gg m$.
Introducing now the  factorization property as an assumption
(see Assumption 1 below) which we assume to be valid
to first order in $\alpha$,  and using a two-time-scale
 perturbation approach, we show that the evolution of the piston
proceeds in two stages with totally different
properties and time scales.
In the first stage, characterized by a time scale $t_1=L\sqrt{Nm/E_0}$
(average time for a particule to collide twice with the piston)
where $E_0$ is the initial energy, the evolution is adiabatic and deterministic,
with properties similar to those obtained in the thermodynamic limit
$M=\infty$ (i.e. $\alpha=0$). In particular, the evolution
can be either weakly or strongly damped and for $M$ sufficiently large
the evolution is independent of $M$ (for fixed $R^{\pm}$).
This evolution proceeds until ``mechanical'' equilibrium 
is reached where the pressures
become equal and it lasts for a time length of order ${\cal O}(M/m)$.

In the second stage, characterized by a time scale 
$t_2=Mt_1/m$, the evolution is strongly dependent on $M$. In fact,
 we shall derive the following scaling property:

\vskip 1mm
\noindent
\underline{\it Main result:} introducing the scaled time variable
$\tau=\alpha t$, the evolution of the piston of mass  
$M\gg m$ is given by:
\be
X_M(\tau)=L\left({1\over 2}-\xi(\tau)\right)\ee
where
\be
{d\xi\over d\tau}=-\;{1\over 3}\;{1\over L}\;\sqrt{E_0\over N}\;
\sqrt{8\over m\pi}\;\;
\left[\;
\sqrt{\displaystyle{N\over N^+}(1+2\xi)}-
\sqrt{\displaystyle{N\over N^-}(1-2\xi)}\;
\right]\ee
Moreover from the knowledge of $\xi(\tau)$, one obtains 
the temperatures and pressures on both sides of the piston, as well
as the temperature of the piston, to lowest significant order in $\alpha$.
In this second stage, 
 the fluctuation-driven evolution is stochastic and proceeds with heat transfer
across the piston towards a state of ``thermal'' equilibrium where $T^-=T^+$.
Investigation to higher orders in $\alpha$ would be necessary
 to conclude that the fluid velocity distributions will ultimately tend 
to Maxwellian distributions.

\vskip 3mm
The same microscopical model was analyzed by Lebowitz, Piasecki and Sinai using
 a different limiting procedure \cite{LPS}.
In their work, they have considered the container to be a cube
of size $L$ and they have taken the limit $L\rightarrow \infty$
with $N^-=N^+\sim L^3$ and $M\sim L^2$.
Using heuristical arguments, they derived autonomous
 coupled equations describing 
the motion of the piston and the fluid  for large $L$.
More recently, exact results were obtained by Chernov, Lebowitz and 
Sinai using this limiting procedure $L\rightarrow\infty$ \cite{chernov}. 
They were able to prove
rigorously that for a time interval sufficiently short so that only the first
and second recollisions of a particle on the piston can occur, 
the random functions
describing the position and the velocity of the piston,
 expressed in terms of scaled variables
$(\tau=t/M,\; Y=X/M)$, converge in probability to some deterministic
functions.
\vskip 3mm
To conclude this introduction, we should mention
that the approach to ``thermal'' equilibrium
was investigated in \cite{munakata}
using an expansion of the master equation for the piston velocity distribution
function in powers of $\sqrt{\alpha}$, as well as numerical simulations, for a gas 
of hard rods.

Another interesting discussion was presented in \cite{white}, where
the power spectrum and the time correlation function of the piston were
calculated at equilibrium for a two-dimensional system of hard disks
by molecular dynamics simulations. From this analysis, one  can
 also deduce a two-time-scale relaxation
towards equilibrium.
The problem of approach to equilibrium and the question of heat transfer for this 2-dimensional
system of hard disks was analyzed in \cite{kestemont} and \cite{vandenbroeck}
using molecular dynamics simulations and a linear Langevin equation.


\section{Equations for the moments $\langle V^n\rangle$ of the piston velocity}
In the spirit of BBGKY hierarchy, we shall characterize
 the velocity distribution 
$\Phi(V,t)$ through its  moments:
\be
\bar{V}(t)=\langle V\rangle (t)=\int_{-\infty}^{\infty} V\Phi(V,t)dV\ee
and \be
\langle V^n\rangle (t)=\int_{-\infty}^{\infty} V^n\Phi(V,t)dV
\ee
Introducing the  notation:
\be
\tilde{F}_k(V,\rho_{surf}^{\pm})=\left[\int_V^{\infty}
(v-V)^{k}\rho_{surf}^-(v,V,t)dv -\int_{-\infty}^V
(v-V)^{k}\rho_{surf}^+(v,V,t)dv\right]
\ee
where $k\geq 0$ and $\rho_{surf}^{\pm}(v,V,t)$
is the two-point correlation function for one molecule
on the left $(-)$, resp. on the right $(+)$, and the piston, 
then by integration of Liouville equation 
 over all variables except the piston velocity $V$, we have obtained 
the following evolution equation for $\Phi$ \cite{GPL}:
\be
{1\over \gamma}\;\partial_t\Phi(V,t)=
\sum_{k=0}^{\infty}\;
{(-1)^{k+1}\alpha^k\over (k+1)!} \;\;
{\partial^{k+1}\tilde{F}_{k+2}\over \partial V^{k+1}}(V,\rho_{surf}^{\pm})
\ee
where $\gamma=A\alpha= 2mA/(M+m)$ and 
with $\partial_t\Phi(V,t)$  defined in the framework of generalized
functions. Hence, {\it by definition}, integration over
$V$ commute with $\partial_t$ and with the infinite summation 
in any average with
respect to $\Phi$:
\be
{1\over \gamma}\;{d\langle h(V)\rangle \over dt}=
\int_{-\infty}^{\infty}\;h(V)\;\partial_t\Phi(V,t)\;dV=
\sum_{k=0}^{\infty}\int_{-\infty}^{\infty}\;h(V)\;{(-1)^{k+1}\alpha^k\over (k+1)!} \;
{\partial^{k+1}\tilde{F}_{k+2}\over \partial V^{k+1}}(V,t)\;dV
\ee
In particular, we have:
\be
{1\over \gamma}\;{d\langle V^n\rangle \over dt}=\sum_{k=0}^{\infty}\int_{-\infty}^{\infty}
V^n\;{(-1)^{k+1}\alpha^k\over (k+1)!} 
{\partial^{k+1}\tilde{F}_{k+2}\over \partial V^{k+1}}(V,\rho_{surf}^{\pm})dV
\ee
Under the hypothesis that $V\rightarrow \rho_{surf}^{\pm}(v,V,t)$ decreases faster
than any power of $V$, uniformly in $v$,
we may integrate by parts, which yields:
$$
{1\over \gamma}\;{d\langle V^n\rangle \over dt}=\sum_{k=0}^{n-1}\;
\int_{-\infty}^{\infty}{n!\;\alpha^k\over (k+1)!} \;{V^{n-k-1}\over
(n-k-1)!}\;\;\tilde{F}_{k+2}(V,\rho_{surf}^{\pm})\;dV$$
\be
+\sum_{k=n}^{\infty}\;\int_{-\infty}^{\infty}{n!\;(-1)^{k+1+n}\alpha^k\over (k+1)!}
\;\;{\partial^{k+1}\tilde{F}_{k+2}\over \partial V^{k+1}}
(V,\rho_{surf}^{\pm})\;dV 
\ee
All terms with $k\geq n$, appearing as pure derivatives with respect to $V$, vanish
after integration on the whole $V$-axis, due to the fast decrease of
$V\rightarrow \rho_{surf}^{\pm}(v,V,t)$ at infinity.
Hence the equation simplifies into:
\be
{1\over \gamma}\;{d\langle V^n\rangle \over dt}=\sum_{k=0}^{n-1}
\int_{-\infty}^{\infty}{n!\;\alpha^k\over (k+1)!} {V^{n-k-1}
\over (n-k-1)!}\tilde{F}_{k+2}(V,\rho_{surf}^{\pm})dV
\ee
where $\tilde{F}_k$ depends functionally on $\rho_{surf}^{\pm}(v,V,t)$.

\section{Factorization property}

In \cite{GPL}, we have shown that in the limit $\alpha=0$, the
two-point correlation functions $\rho_{surf}^{\pm}(v,V,t)$
have the factorization property.
Although we can not prove it, we expect that 
in the limit  $\alpha\rightarrow 0$ which we now consider,
and for initial conditions such that the evolution is smooth,
this property will still be valid to first order
${\cal O}(\alpha)$ where the leading behavior is $\Phi(V,t)=
\delta(V-\bar{V}(t))$.
Therefore in the following perturbation approach, we shall
 consider that the following assumption holds to first order in
$\alpha$:

\vskip 1mm
\noindent
\underline{\it Assumption 1 (factorization property):}
the two-point correlation functions have the following 
factorization property at first order in
$\alpha$:
\be 
\rho_{surf}^{\pm}(v,V,t)=\rho_{surf}^{\pm}(v,t)\;\Phi(V,t)\ee
\vskip 1mm
\noindent
Under this factorization property, 
we have $\widetilde{F}_k=F_k\Phi$ where:
\be
F_k(V,\rho_{surf}^{\pm})=F_k^-(V,\rho_{surf}^{-})-F_k^+(V,\rho_{surf}^{+})\;=\;
\int_V^{\infty}
(v-V)^{k}\rho_{surf}^-(v,t)dv -\int_{-\infty}^V
(v-V)^{k}\rho_{surf}^+(v,t)dv
\ee
and the evolution of the moments $\langle V^n\rangle$
of the piston velocity satisfy the equation:
\be\label{10}
{1\over \gamma}\;{d\langle V^n\rangle \over dt}=\sum_{k=0}^{n-1}
\int_{-\infty}^{\infty}{n!\;\alpha^k\over (k+1)!} {V^{n-k-1}
\over (n-k-1)!}\;F_{k+2}(V,\rho_{surf}^{\pm})\;\Phi(V,t)\;dV
\ee
Similarly, under the factorization assumption,
the distributions  $\rho^{\pm}(x,v,t)$
of the fluid molecules satisfy
the Boltzmann equations with boundaries \cite{GPL}:

\vskip 1mm
$(\partial_t+v\partial_x)\rho^-(x,v,t)=\delta (x+L)v\rho^-(-L,v,t)$
\be\label{boltz1}
+\delta(x-X(t))[V(t)-v][\theta(V(t)-v)
\rho^-(X(t), v^{\prime},t)+\theta(v-V(t))\rho^-(X(t), v,t)]
\ee
\vskip 3mm $
(\partial_t+v\partial_x)\rho^+(x,v,t)=-\delta (x-L)v\rho^+(L,v,t)$
\be\label{boltz2}
-\delta(x-X(t))[V(t)-v][\theta(v-V(t))
\rho^+(X(t), v^{\prime},t)+\theta(V(t)-v)\rho^+(X(t), v,t)]
\ee
where $v^{\prime}=2V-v+\alpha(v-V)$
is the velocity of the molecules after their collision 
onto the piston.
Note that $\rho^{\pm}(X(t), v,t)$ is what we denote
 $\rho^{\pm}_{surf}(v,t)$.

\vskip 3mm
It is straighforward to check that:
\be\label{15}
{dF^{\pm}_k\over dV}(V,\rho_{surf}^{\pm} )
=-\,k\,F^{\pm}_{k-1}(V,\rho_{surf}^{\pm} )
\hskip 15mm ({\rm if}\;\;k\geq 1)\ee
which leads to define $F^{\pm}_k$ for $k<0$ by:
\be
\left({d\over dV}\right)^rF_0^{\pm}
\doteq F^{\pm}_{-r}=\pm\;\left({d\over dV}\right)^{r-1}\;
\rho_{surf}^{\pm}(V,t)\hskip 15mm (r\geq 1)\ee
We define the densities $\rho_{surf}^{\pm}$
(not to be confused with the distributions
$\rho_{surf}^{\pm}(v)$),
the temperatures $T_{surf}^{\pm}$ and the pressures
$p_{surf}^{\pm}$ at the surface of the piston, on each side:
\begin{eqnarray}
\rho_{surf}^-&\doteq& 2\int_0^{\infty}\rho_{surf}^{-}(v,t)dv
=2F_0^-(V=0,\rho_{surf}^{-})\\
&&\nonumber\\
\rho_{surf}^+&\doteq& 2\int_{-\infty}^{0}\rho_{surf}^{+}(v,t)dv
=2F_0^+(V=0,\rho_{surf}^{+})\\
&&\nonumber\\
p_{surf}^{\pm}&\doteq &2mF_2^{\pm}(V=0)\doteq 
\rho_{surf}^{\pm}k_BT_{surf}^{\pm}
\end{eqnarray}
It is obvious from the definition that $F_2^-(V)$
and $-\,F_2^+(V)$ are decreasing functions of $V$,
which allows to express $F_2^{\pm}(V)$ and  $F_0^{\pm}(V)$ in the following form:
\begin{eqnarray}
2mF_2^{\pm}(V)&\doteq &p_{surf}^{\pm}\;\pm\;
\left({M+m\over A}\right)\;\lambda^{\pm}(V)V\\
&&\nonumber\\
2F_0^{\pm}(V)&\doteq &\rho_{surf}^{\pm}\;\pm\;
\left({M+m\over A}\right)\;\left({\widetilde{\lambda}^{\pm}(V)
\over k_BT_{surf}^{\pm}}\right)\;V
\end{eqnarray}
and $\lambda^{\pm}(V)$ have the physical meaning
of  friction coefficients.
Denoting $\lambda(V)=\lambda^{-}(V)+\lambda^{+}(V)$ yields:
\be
2mF_2(V)=(p^--p^+)\;-\;\left({M+m\over A}\right)\;\lambda(V)V\ee
From Eq. (\ref{15}), we have also $2mF_2^{\pm}(V)=p^{\pm}-4mF_1^{\pm}(0)V+{\cal O}(V^2)$, so that,
denoting simply $\lambda^{\pm}=\lambda^{\pm}(V=0)$:
\be
F_1^{\pm}(V=0)\;=\;\mp\;{1\over 4m}\;\left({M+m\over A}\right)\;\lambda^{\pm}\ee

\section{Irreducible moments}

We denote $F_n^{(r)}$ the $r$-th derivative of $F_n$
(with respect to $V$); from Eq. (\ref{15}),  these $F_n^{(r)}$ 
are functional of $\rho^{\pm}_{surf}(v,t)$, proportional
to $F_{n-r}(V, \rho^{\pm}_{surf})$.
We thus have:
\be\label{21}
F_{n}(V, \rho^{\pm}_{surf})=\sum_{r=0}^{\infty} {1\over r!}\;
F_{n}^r(\bar{V}, \rho^{\pm}_{surf})(V-\bar{V})^r\ee
where:
\begin{eqnarray}
(F_n^{\pm})^{(r)}&=&
(-1)^r\;{n!\over (n-r)!}\;F_{n-r}^{\pm} \hskip 15mm{\rm for}\;\;\;r\leq n\\
&&\nonumber\\
(F_n^{\pm})^{(r)}&=&\pm\;(-1)^n\;n!
\left({d\over dV}\right)^{r-n-1}\rho^{\pm}_{surf}(V,t)
\hskip 15mm{\rm for}\;\;\;r\geq n+1
\end{eqnarray}
Plugging the expansion Eq. (\ref{21}) in Eq. (\ref{10}) yields for 
$n=1$:
\be\label{25}
{1\over\gamma} {d\bar{V}\over dt}=
F_2(\bar{V},t)+\sum_{r\geq 2}{\Delta_r\over r!}\;
F_2^{(r)}(\bar{V},t)
\ee
where:
\be
F_2^{(0)}=F_2\hskip 8mm
F_2^{(1)}=-2F_1\hskip 8mm
F_2^{(2)}=2F_0\hskip 8mm
F_2^{(3)}=2F_{-1}=
-2[\rho_{surf}^-(V,t)-\rho_{surf}^+(V,t)
]\ee
and $\Delta_r$ are the irreducible moments:
\be\label{11}
\Delta_r\equiv \int_{-\infty}^{\infty}(V-\bar{V}(t))^r\Phi(V,t)dV
=\sum_{q=0}^r (-1)^q\;{r!\over q! (r-q)!}
\; \bar{V}(t)^q\;\langle V^{r-q}\rangle \ee
with
\be \bar{V}(t)=\int_{-\infty}^{\infty}V\Phi(V,t)dV\ee
 the average velocity of the piston. Let us note that
\begin{eqnarray}
\Delta_0&=&1\\
\Delta_1&=&0\\
\Delta_2&=&\langle V^2\rangle -\bar{V}^2(t)
\end{eqnarray}
We can invert Eq. (\ref{11}) to express the moments
$\langle V^s\rangle $ as functions of the irreducible moments
(cumulant expansion):
\be\label{32}
\langle V^s\rangle =
\bar{V}^s\;+\;\sum_{q=2}^s\;\;{s!\over q!(s-q)!}\;
\bar{V}^{s-q}\;\Delta_q\ee
Finally replacing  the function $F_{k+2}(V,\rho_{surf}^{\pm})$ 
by its expansion in powers of $(V-\bar{V}(t))$ 
around the average velocity $\bar{V}(t)$
and using Eq. (\ref{32}) lead 
to the following evolution equations for the irreducible
moments ($s\geq 2$):
$$
{1\over\gamma} {d\Delta_s\over dt}=
-s\Delta_{s-1}\left[ \sum_{r\geq 2} {1\over r!}
\Delta_{r}F_2^{(r)}\right]
-2s\left[\sum_{n\geq 0}{1\over (n+1)!}
\Delta_{s+n}F_1^{(n)}\right]
$$
\be \label{18}
+\alpha \;\left[\sum_{k=0}^{s-2}\alpha^k\;{s!\over (k+2)!(s-2-k)!}
\left[\sum_{n\geq 0}{1\over n!}\Delta_{s-2-k+n}
 F_{3+k}^{(n)}\right]\right]
\ee
where we recall that
$F_k$  are functions of $\bar{V}$ and  functional
of $\rho^{\pm}(.,t)$.
%


For  $s=2$, it comes:
\be
{1\over\gamma} {d\Delta_2\over dt}=
-4\sum_{n\geq 0}\;{\Delta_{n+2}\over (n+1)!}\;F_1^{(n)}\;
+\;\alpha \;\sum_{n\geq 0}\;{1\over n!}\;\Delta_n\;F_3^{(n)}
\ee
Using the fact that  $\Delta_1=0$, 
$F_{1}^{(n)}=-{1\over 2}F_2^{(n-1)}$
and $F_3^{(r)}=-3F_2^{(r-1)}$ if $r\geq 1$, we may rewrite this
equation:
\be\label{35}
{1\over\gamma} {d\Delta_2\over dt}=
\sum_{r\geq 2}\;{\Delta_r\over r!}\;(2r-3\alpha)\;F_2^{(r-1)}
\;+\;\alpha F_3
\ee

\section{Evolution at first order in $\alpha$}
A qualitative analysis of Eq. (\ref{18}) shows that:
\be\label{36}
\Delta_s\sim \;\alpha^{[{(s+1)\over 2}]}\ee
where $[(s+1)/2]$ is the integral part of $(s+1)/2$.
Since we want to restrict our study to first order in 
$\alpha$, assuming that Eq. (\ref{36}) is valid, 
we only have to consider
the evolution of $\bar{V}$ 
 and  $\Delta_2$, Eqs. (\ref{25}) and (\ref{35}),
which we supplement with the equations for the energies
$ \langle E^{\pm}\rangle $ of the gas in the left and right compartments,
restricted similarly to first order in $\alpha$.
We thus obtain to first order in $\alpha$ the following set  
 of coupled
{\it deterministic} equations:
\begin{eqnarray}
\hskip 10mm \displaystyle
{1\over\gamma} {d\bar{V}\over dt}&=& F_2+
\Delta_2F_0\label{40a}\\
&&\nonumber\\
\hskip 10mm\displaystyle
{1\over\gamma} {d\Delta_2\over dt}&=&
-4\Delta_2F_1+
\alpha F_3\hskip 10mm\label{40b}
\\&&\nonumber\\
\displaystyle\hskip 10mm
{1\over\gamma} {d\langle E^-\rangle 
\over dt}&=&-\,M\bar{V}[F_2^-+\Delta_2F_0^-]
+{M\over 2}[4\Delta_2F_1^--\alpha F_3^-]\hskip 10mm \label{40c}\\
&&\nonumber\\
\displaystyle
{1\over\gamma} {d\langle E^+\rangle \over dt}&=& M\bar{V}[F_2^++\Delta_2F_0^+]
-{M\over 2}[4\Delta_2F_1^+-\alpha F_3^+]\hskip 10mm\label{40d}
\end{eqnarray}
\vskip 4mm
\noindent
and we recall that all the functions $F$  are functions of 
$\bar{V}(t)$ and functionals of $\rho^{\pm}_{surf}(v,t)$. 
We should  note that the set of Eqs. (\ref{40a})-(\ref{40d}) have a constant
of motion, i.e.
\be\label{Eint}
\langle E^-\rangle +\langle E^+\rangle +{1\over 2}\;M(\bar{V}^2+\Delta_2)=E_0=const.
\ee
which reflects the conservation of energy at the microscopic level.

\vskip 3mm\noindent
{\bf Remark 1:} \ The evolution is described by Eqs. (\ref{40a})
and (\ref{40b}) together with Boltzmann equations for the fluid
Eqs. (\ref{boltz1}) and (\ref{boltz2}).
In the next section, we shall introduce the 
``average assumption'' which will then enable us to ignore the equation
for the fluids and to replace them by Eqs. (\ref{40c}) and (\ref{40d}).

\vskip 3mm\noindent
{\bf Remark 2:} \ Recalling that $\Delta_2(t=0)=0$,
it is straighforward using Gronwall lemma to check the consistency
of the scaling  hypothesis, i.e.  $\Delta_2={\cal O}(\alpha)$, 
from the above evolution equation (\ref{40b}).
Indeed, $F_3$ remains of order ${\cal O}(1)$  (upper bound 
denoted $\sup(F_3)$) 
and $F_1$ is strictly positive since $F_1=0$ would imply that:
\be
\left\{
\begin{array}{lr}
\rho^{-}_{surf}(v,t)=0&\hskip 10mm\forall v\geq \bar{V}(t)\\
\rho^{+}_{surf}(v,t)=0&\hskip 10mm\forall v\leq \bar{V}(t)
\end{array}
 \right.
\ee
which is precluded by the relations:
\be
\left\{
\begin{array}{lr}
\forall v\geq \bar{V}(t)\hskip 10mm&(1-\alpha)\,\rho^{-}_{surf}(v,t)=
\rho^{-}_{surf}(2\bar{V}_t-v+\alpha(v-\bar{V}_t),t)\\
\forall v\leq \bar{V}(t)\hskip 10mm&(1-\alpha)\,\rho^{+}_{surf}(v,t)=
\rho^{+}_{surf}(2\bar{V}_t-v+\alpha(v-\bar{V}_t),t)
\end{array}
 \right.
\ee
Therefore
 $\Delta_2$ is bounded by $\alpha\sup(F_3)/\inf(F_1)$
i.e. $\Delta_2={\cal O}(\alpha)$.

\vskip 3mm\noindent
{\bf Remark 3:}  \ In  our  first paper \cite{GPL},
we have considered the case $\alpha=0$,
i.e. the thermodynamic limit for the piston,
 and discussed the evolution
described by $\Delta_2=0$ and:
\begin{eqnarray}
{1\over\gamma}\;{dV\over dt}&=&F_2(V)
\\
&&\nonumber\\
{1\over\gamma}\;{d\langle E^-\rangle \over dt}&=& 
-MVF_2^-(V)\\
&&\nonumber\\
{1\over\gamma}\;{d\langle E^+\rangle\over dt} &=&
 MVF_2^+(V)
\end{eqnarray}

\section{Adiabatic evolution for short time}

We want to investigate the evolution of the piston under
the initial condition where $X(t=0)=X_0$, $V(t=0)=0$
and the fluids on both sides of the piston are in equilibrium, described
by Maxwellian distribution of velocity, with temperatures 
$T_0^{\pm}$. The initial conditions are such that 
$|T_0^+-T_0^-|={\cal O}(1)$ and 
$|p_0^+-p_0^-|={\cal O}(1)$. Since $\alpha \ll 1$, the
perturbation approach shows that 
in a first stage, we can restrict our analysis to order
zero in $\alpha$ and thus we recover the results of our previous
paper (where $\alpha=0$), except that they will now be valid
only for finite time.
In this first stage, the motion of the piston is a {\it deterministic}
(no velocity fluctuations,
$\Delta_2=0$)
 {\it adiabatic} (no heat transfer between the compartments)
evolution towards
mechanical equilibrium. 
This first stage ends when $p^--p^+={\cal O}(\alpha)$
(but  $T^--T^+$ is still ${\cal O}(1)$).
At this 
point, the term $F_2$ which appears in the evolution of
$\bar{V}$, Eq. (\ref{40a}), becomes of order $\alpha$ and
the first order terms
of the perturbation approach must now  be taken into account.
Let us note that if we introduce the ``average assumption''
(see below, Assumption 2), the state at the end of this first stage is given by the final
state derived in \cite{GPL} for $\alpha=0$:
\begin{eqnarray}
p_{ad}^{\pm}&=&p_0+{\cal O}(\alpha)\label{50}\\
&&\nonumber\\
T_{ad}^-&=&\left({N\over N^-}\right)\;T_0\;{X_{ad}\over L}+{\cal O}(\alpha)\\
&&\nonumber\\
T_{ad}^+&=&\left({N\over N^+}\right)\;T_0\;
\left(1-{X_{ad}\over L}\right)+{\cal O}(\alpha)
\end{eqnarray}
where:
\be
k_BT_0=\left({AL\over N}\right)\;p_0={2E_0\over N}\;,
\hskip 15mm 
N=N^-+N^+\;,\ee
and $X_{ad}$ is the solution of 
\be\label{Xad}
\sqrt{\left({N\over N^-}\right)X_{ad}^3}
\;-\;\sqrt{\left({N\over N^+}\right)(L-X_{ad})^3}
\;=\;\sqrt{L\over T_0}\;C\ee
where the constant $C$ is related to the initial conditions according to
\be
C= 
\sqrt{T^-(0)}X_0-\sqrt{T^+(0)}(L-X_0)
\ee

\section{Slow relaxation towards thermal equilibrium}
           
The perturbation approach developped here (at order 1 in $\alpha$)
intends to reach the long-time behavior of the piston motion
in the case where $M\gg m$. 
We claim that the  fluctuations of the piston velocity $V$
enter the scene only in a second stage, of time scale ${\cal O}(1/\alpha)$,
once the pressure difference has become of order ${\cal O}(\alpha)$.
In order to prove this assertion, we
have to develop a boundary-layer-type
perturbation approach\cite{nayfeh};
indeed, at the end of the fast relaxation
towards mechanical equilibrium, the standard perturbation approach
(with  time  variable $t$) becomes singular and fails to
give access to the further evolution of the system.
 The relevant time variable to be used
in order to reach the second stage of the evolution is the rescaled variable:
\be
\tau\doteq \alpha\;t\ee
When the evolution is described in terms of this rescaled time, 
the first stage collapses into a boundary layer ($\tau\ll 1$)
whereas the focus now bears on the second (previously asymptotic) stage. 
We  introduce a rescaled velocity $\widetilde{V}(\tau)$,
describing a slow motion of the piston:
\be
\bar{V}(t)\doteq \alpha\;\widetilde{V}(\tau)
\hskip 10mm\mbox{\rm with}
\hskip 3mm \widetilde{V}(\tau)={dX\over d\tau}\ee
The motion is now  driven by fluctuations and the velocity fluctuations
play a crucial role; they allow to  define the piston temperature $T^P$:
\be\label{58}
\Delta_2(t)\doteq \alpha\;\widetilde{\Delta_2}(\tau)
\hskip 10mm\mbox{\rm with}
\hskip 3mm \widetilde{\Delta_2}\doteq {k_BT^P\over 2m}\ee
Finally, mechanical equilibrium has been reached in the first
stage of the evolution and only fluctuations around  
mechanical equilibrium are to be observed in the second stage.
We thus introduce:
\be
(p^--p^+)(t)\doteq \alpha\;\widetilde{\Pi}(\tau)
\ee
In this second stage, the rescaled quantities are of order 1 and this stage
ends  when they become of order $\alpha$. At this point, we should then
consider the corresponding equation to order $\alpha^2$.
We now investigate the consequences  of the perturbation approach
to order 1 in $\alpha$.
Equations (\ref{40a})-(\ref{40d})
can be written in terms of the rescaled quantities:
\be
\left\{
\begin{array}{l} \displaystyle
{\alpha\over\gamma}\;{d \widetilde{V}\over d\tau} \;=\;
{\widetilde{\Pi}\over 2m}-2F_1\widetilde{V}+\widetilde{\Delta_2}F_0
\label{eqVtilde}\\
\\ \displaystyle
{\alpha\over\gamma}\;{d \widetilde{\Delta_2}\over d\tau}\;=\;
-4\widetilde{\Delta_2}F_1+F_3
\label{eqDeltatilde}
\end{array}
\right.
\ee
Another set of  equations describes the evolution of the gas 
energies
(at lowest order in $\alpha$):
\be
\left\{
\begin{array}{l}\displaystyle
{1\over N^-}{d\langle E^-\rangle\over d\tau}\;=
\;-\,2m\left({A\over N^-}\right)
\;\widetilde{V}\;[F_2^-+\Delta_2F_0^-]
\;+\;m\left({A\over N^-}\right)
[4\widetilde{\Delta_2}F_1^--F_3^-]\label{13}
\\
\\
\displaystyle
{1\over N^+}{d\langle E^+\rangle\over d\tau}\;=\;2m
\left({A\over N^+}\right)
\;\widetilde{V}\;[F_2^++\Delta_2F_0^+]
\;-\;m\left({A\over N^-}\right)
[4\widetilde{\Delta_2}F_1^+-F_3^+]\label{14}
\end{array}
\right.
\ee
Consistency of the perturbation approach
then requires to take the value of all the functions $F_j$
at $\bar{V}=0$ (as soon as $\bar{V}$ remains of order ${\cal O}(\alpha)$,
i.e. $\widetilde{V}={\cal O}(1)$), and therefore to  replace
Eqs. (\ref{13})  by:
\be\label{E}
\left\{
\begin{array}{l}\displaystyle
{1\over N^-}{d\langle E^-\rangle\over d\tau}\;=
\;-\,\left({A\over N^-}\right)
\;\widetilde{V}\;p_{surf}^-
\;+\;m\left({A\over N^-}\right)
[4\widetilde{\Delta_2}F_1^--F_3^-]
\\
\\
\displaystyle
{1\over N^+}{d\langle E^+\rangle\over d\tau}\;=\;2m
\left({A\over N^+}\right)\;\widetilde{V}\;p_{surf}^+
\;-\;m\left({A\over N^-}\right)
[4\widetilde{\Delta_2}F_1^+-F_3^+]
\end{array}
\right.
\ee

\section{Slaving principle (consistency condition)}
Equations (\ref{eqVtilde}) show that
$ \widetilde{V}$ and $\widetilde{\Delta_2}$ are slaved to the slow relaxation of
the gases towards thermal equilibrium appearing in
the $\rho^{\pm}$-dependence of the $F_j$:
\vskip 4mm
\be\label{cons1}
\begin{array}{rcl}
\hskip 10mm \displaystyle
{\widetilde{\Pi}\over 2m}&=&2\widetilde{V}F_1(0)
-\displaystyle{F_3(0)F_0(0)\over 4F_1(0)}
\\
&&\nonumber\\\displaystyle
\widetilde{\Delta_2}&=&\displaystyle{F_3(0)\over 4F_1(0)}
\hskip 10mm 
\end{array}
\ee
\vskip 4mm
\noindent
We call \underline{\it consistency condition}
such a lower-order resolution of evolution equations
of the general  form $\alpha\; dA/d\tau=r.h.s.(\tau)={\cal O}(1)$, leading
 to the instantaneous relation $r.h.s. (\tau)=0$, up to terms of order 
${\cal O}(\alpha)$.  Indeed, it merely follows from a term-wise identification 
of the expansion in powers of $\alpha$, given that $A$ actually 
is a non trivial 
function of $\tau$ when $\alpha\rightarrow 0$\cite{haken}.
It can also be termed a quasi-steady-state approximation
(as for instance in the context of enzymatic catalysis \cite{murray}).
We shall make further encounter of this argument.
As mentionned above, the consistency of the lower order 
perturbation approach also requires to set $V=0$ in
 functions $F_j$.

\section{Average assumption}
At this stage, to simplify our analysis, we introduce the following:

\vskip 1mm
\noindent
\underline{Assumption 2 (average assumption):}
The temperatures and the densities at the surface of
 piston coincide at order 1 in $\alpha$ with the average
energy and density in the fluids in the respectively left/right
compartments:
\vskip 1mm
\noindent
a) \ $T_{surf}^{\pm}=T^{\pm}$ \hskip 3mm where  \hskip 3mm
$N^{\pm}\,k_B\,T^{\pm}=2\langle E^{\pm}\rangle $.

\vskip 1mm
\noindent
b) \ $\rho_{surf}^{\pm}=\rho^{\pm} $ \hskip 3mm
where  \hskip 3mm$\displaystyle\rho^-\;=\;{N^-\over A\,X}$, \hskip 3mm
\  $\displaystyle\rho^+\;=\;{N^+\over A\,(L-X)}$,
\hskip 3mm  and then \hskip 3mm $p_{surf}^{\pm}=p^{\pm}
=\rho^{\pm}k_BT^{\pm}$. 

\vskip 3mm
From Eq. ( \ref{Eint}), we get:
\be
N^-k_BT^-\;+\;N^+k_BT^+=2E_0-M(\Delta_2+\bar{V}^2)\ee
or equivalently
\be
Xp^-+(L-X)p^+={2E_0\over A}-{M\over A}(\Delta_2+\bar{V}^2)
\ee
Therefore:
\begin{eqnarray}
p^-&=&p_0\;+\;\alpha\;\left[\left(1-{X\over L}\right)\widetilde{\Pi}
-{M\over AL}(\widetilde{\Delta_2}+\alpha\widetilde{V}^2)
\right]\label{eqp-}\\
&&\nonumber\\
p^+&=&p_0\;-\;\alpha\;\left[\left({X\over L}\right)\widetilde{\Pi}
+{M\over AL}(\widetilde{\Delta_2})+\alpha\widetilde{V}^2)
\right]\label{eqp+}
\end{eqnarray}
where  by definition $p_0=2E_0/AL$, and thus:
\be
{dp^{\pm}\over d\tau}={\cal O}(\alpha)\ee
Moreover, from Eqs. (\ref{E}) we have:
\be
\left\{
\begin{array}{l}\displaystyle
k_B {dT^-\over d\tau}\;=\;-\,2\left({A\over N^-}\right)
\;\widetilde{V}p^-+
\,2m\left({A\over N^-}\right)
[4\widetilde{\Delta_2}F_1^-+F_3^-]
\;+\;{\cal O}(\alpha)\label{eqT-tilde}
\\
\\
\displaystyle
k_B {dT^+\over d\tau}\;=\;2\left({A\over N^+}\right)
\;\widetilde{V}p^+-\;2m\left({A\over N^+}\right)
[4\widetilde{\Delta_2}F_1^++F_3^+]
\;+\;{\cal O}(\alpha)\label{eqT+tilde}
\end{array}
\right.
\ee
On the other hand, from Assumption 2b  (average assumption),
 we have:
\be\label{72}
p^-={N^-k_BT^-\over AX}
\hskip 20mm
p^+={N^+k_BT^+\over A(L-X)}\ee
which together with Eqs. (\ref{eqT-tilde}) and (\ref{72}),
yields:
\begin{eqnarray}
\left({N^-\over A}\right)\;{1\over p^-}\;{dp^-\over d\tau}
&=& -3\rho^-\widetilde{V}\;+\;
{2m\over k_BT^-}[4\widetilde{\Delta_2}F_1^--F_3^-]\;+
\;{\cal O}(\alpha)\label{eqdp-}\\
&&\nonumber\\
\left({N^+\over A}\right)\;{1\over p^+}\;{dp^+\over d\tau}&=&
3\rho^+\widetilde{V}\;-\;
{2m\over k_BT^+}[4\widetilde{\Delta_2}F_1^+-F_3^+]\;+\;
{\cal O}(\alpha)\label{eqdp+}
\end{eqnarray}
From Eq. (\ref{cons1}) follows that:
\be\label{29}
4\widetilde{\Delta_2}F_1^--F_3^-=
4\widetilde{\Delta_2}F_1^+-F_3^+=
{1\over F_1}\;(F_3^-F_1^+-F_3^+F_1^-)\ee
On the other hand, from Eqs. (\ref{eqp-}) and (\ref{eqp+}),
we have:
\be
\rho^-k_BT^-=\rho^+k_BT^++{\cal O}(\alpha)\ee
Therefore, since $dp^{\pm}/d\tau$ is of order $\alpha$, the consistency
condition of the perturbation approach yields from 
Eqs. (\ref{eqdp-}), (\ref{eqdp+}) and (\ref{29})
the final relation:
\be\label{30}
\left\{
\begin{array}{rcl}
\widetilde{V}&=&\displaystyle {m\over 3}\left({AL\over E_0}\right)
\left({F_3^-F_1^+-F_3^+F_1^-\over F_1}\right)+{\cal O}(\alpha)\\
&&\nonumber\\
\displaystyle{\widetilde{\Pi}\over 2m}&=&\displaystyle{2m\over 3}
\left(\displaystyle{AL\over E_0}\right)
(F_3^-F_1^+-F_3^+F_1^-)\;-\;\displaystyle {F_3F_1\over 4F_1}
+{\cal O}(\alpha)\\ && \\
\widetilde{\Delta_2}&=&\displaystyle {F_3\over 4F_1}
+{\cal O}(\alpha)
\end{array}
\right.\ee

\section{Dimensionless variables}
As usual in perturbation methods, it is convenient to 
introduce dimensionless variables. From
Eq. (\ref{10}) and Assumption 2a, we have:
\be\label{76}
N^-k_BT^-\;+\;N^+k_BT^+=2E_0-M(\Delta_2+\bar{V}^2)
\ee
Let us define:
\be
k_BT_0={2E_0\over N}=\left({AL\over N}\right)p_0
\ee
It is natural to introduce the dimensionless variable $\xi$ 
defined by:
\be\label{37}
N^{\pm}k_BT^{\pm}\doteq
{1\over 2}(1\pm 2\xi)\;\left[Nk_BT_0-M\alpha
(\widetilde{\Delta}_2+\alpha\widetilde{V}^2)
\right]\ee
We thus have from Eqs. (\ref{eqp-}) and (\ref{eqp+}):
\begin{eqnarray}
N^+k_BT^+-N^-k_BT^-&=&
2\xi[ALp_0-M\alpha(\widetilde{\Delta}_2+\alpha\widetilde{V}^2)]\\
&=&p^+A(L-X)-p^-AX\\
&=&p_0AL-\alpha AL
\left[ {X\over L}\widetilde{\Pi}+{M\over AL}
(\widetilde{\Delta}_2+\alpha\widetilde{V}^2)  
\right]\\
&&
\hskip 5mm-2p_0AX
+\alpha AX\left[\left({2X\over L}-1\right)\widetilde{\Pi}
+{2M\over AL}
(\widetilde{\Delta}_2+\alpha\widetilde{V}^2)  
\right]
\end{eqnarray}
Therefore:
\be
2\xi\left[1-\alpha\;{M\over ALp_0}\widetilde{\Delta}_2\right]=
\left(1-{2X\over L}\right)+{\alpha\over p_0}
\left[\left({X\over L-1}\right)
{M\over AL}\;\widetilde{\Pi}\;
\widetilde{\Delta}_2
\right]+
{\cal O}(\alpha^2)\ee
The variable $\xi$ remains ${\cal O}(1)$ in the second stage
of the evolution, hence the consistency of
the perturbation approach requires to truncate
the previous equation at lower order in $\alpha$,
which yields:
\be\label{38} 
\xi={1\over 2}-{X\over L}+{\cal O}(\alpha)\ee
According to this  the consistency  condition, we have:
\be
{d\xi\over d\tau}=-\;{\widetilde{V}\over L}\ee
and from Eq. (\ref{30}), we obtain our final equation:
\be \label{84}
{d\xi\over d\tau}=-\; {m\over 3}\;
\left(A\over E_0\right)\;
{1\over F_1}\;(F_3^-F_1^+-F_3^+F_1^-)\ee
\section{Maxwellian identities}
Let us introduce at this point the following:
\vskip 1mm\noindent
\underline{Assumption 3 (Maxwellian identities):} 
\vskip 1mm
\noindent
a)
The relation between
the functionals $F_1$ and $F_3$ obtained when
$\rho^{\pm}$ are Maxwellian 
remains valid  here at order $\alpha$, i.e.
\be
F_3^{\pm}(V)={2k_BT^{\pm}\over m}\;
F_1^{\pm}(V)
\;-\;V\,F_2^{\pm}(V)
+{\cal O}(\alpha)\hskip 7mm{\rm provided }
\hskip 3mm V={\cal O}(\alpha)
\ee

\vskip 1mm
\noindent
b)  $F_1^{\pm}=F_1^{\pm}(V=0)$
coincide at order $\alpha$ with the value
given by Maxwellian distributions, i.e.:
\be
F_1^{\pm}(0)=\mp\;\rho^{\pm}\;\sqrt{k_BT^{\pm}\over 2m\pi}
+{\cal O}(\alpha)\ee

\vskip 3mm
From Assumption 3a, we  have:
\be 
F_3^-F_1^+-F_3^+F_1^-={2k_B\over m}\;
(T^--T^+)F_1^-F_1^+\ee
and thus:
\begin{eqnarray}
{d\xi\over d\tau}&=&-\,k_B(T^--T^+)\left({2A\over 3E_0}\right)
{F_1^-F_1^+\over F_1}\\
&&\nonumber\\
\widetilde{V}&=&k_B(T^--T^+)\left({2AL\over 3E_0}\right)
{F_1^-F_1^+\over F_1}\label{86}\\
&&\nonumber\\
\widetilde{\Delta}_2&=&{k_B\over 2m}\;
{(T^-F_1^--T^+F_1^+)\over F_1}\\
&&\nonumber\\
{\widetilde{\Pi}\over 2m}&=&
k_B(T^--T^+)\left[
\left({4AL\over 3E_0}\right)F_1^-F_1^+
+\left({E_0\over AL}\right)\;{\widetilde{\Delta}_2\over k_B^2T^-T^+}
\right]\label{88}
\end{eqnarray}
We remark that whatever  are the distributions
$\rho^{\pm}(v,t)$, then as long as Assumption 3a is satisfied, we have
$\widetilde{V}>0$ if and only if $T^+>T^-$, i.e. the piston moves in the direction
of the warmer side, which
is typical of heat transfer since $p^+=p^-+
{\cal O}(\alpha)$.
\vskip 3mm
To obtain an explicit equation for the evolution, we shall now use the 
Assumption 3b above, together with the fact that 
$\rho^{\pm}k_BT^{\pm}=p_0+{\cal O}(\alpha)$.
We thus have:
\be\label{*}
{F_1^-F_1^+\over F_1}=\-\;\sqrt{2\over m\pi}\;
\left({E_0\over AL}
\right)\; {1\over \sqrt{k_B}}\;
{1\over \sqrt{T^-}+\sqrt{T^+}}\ee
and Eq. (\ref{84}) leads to the main result of our paper:
\be\label{93}
{d\xi\over d\tau}=-\;{1\over 3}\;{1\over L}\;\sqrt{E_0\over N}\;
\sqrt{8\over m\pi}\;\;
\left[\;
\sqrt{\displaystyle{N\over N^+}(1+2\xi)}-
\sqrt{\displaystyle{N\over N^-}(1-2\xi)}\;
\right]
\ee
Finally introducing the dimensionless variable $s$ by:
\be
s=\tau\;{2\over 3L}\;\displaystyle
\sqrt{k_B\over m\pi}\;\displaystyle\sqrt{{2N^-\over N}\;T_0^-
+{2N^+\over N}\;T_0^+}\ee
we have:

\vskip 1mm
\noindent
\underline{\it Main result:} under Assumptions 1, 2 and 3, 
the evolution of the system 
in the second stage is described in terms of the dimensionless variables by
\be\label{main}
{d\xi\over ds}=-\;\left[
\sqrt{{N\over 2N^+}(1+2\xi)}\;
-\;\sqrt{{N\over 2N^-}(1-2\xi)}\;
\right]\ee
together with the ``initial condition'' (in fact the matching condition
of the boundary-layer-type perturbation approach)
which is from Eq. (\ref{38}):
\be\label{94}
\xi(s=0)\;=\;{1\over 2}-{X_{ad}\over L}
\ee
with $X_{ad}$ being the position of the piston
at the end of the first stage, Eq. (\ref{Xad}), i.e. 
at mechanical equilibrium.
Integrating Eq. (\ref{93}) with Eq. (\ref{94}) yields the evolution
$\xi=\xi(s)$ from which we obtain the position and the temperatures:
\begin{eqnarray}
X&=&L\left({1\over 2}-\xi\right)\\
&&\nonumber\\
T^{\pm}&=&(1\pm 2\xi)\;\;{N^-T_0^-+N^+T_0^+\over 2N^{\pm}} \label{100}
\end{eqnarray}
together with $\widetilde{V}$, $\widetilde{\Delta_2}$
and $\widetilde{\Pi}$ obtained from Eqs. (\ref{86})-(\ref{88})
and Eq. (\ref{*}), i.e.:
\be\label{92}
\left\{
\begin{array}{l}
\widetilde{V}= \displaystyle
{1\over 3}\displaystyle
\sqrt{8k_B\over \pi m}\;
\left(\sqrt{T^+}-\sqrt{T^-}\right)\\
\\
\widetilde{\Delta_2}=\displaystyle{k_B\over 2m}\sqrt{T^-T^+}
\hskip 15mm {\rm i.e.}\;\;T^P=\sqrt{T^+T^-}\label{101}\\
\\
\widetilde{\Pi}=
\left(\displaystyle{E_0\over AL}\right)\;\left(\displaystyle{T^+-T^-\over \sqrt{T^-T^+}}\right)
\left(\displaystyle{16\over 3\pi}-1\right)
\end{array}
\right.\ee
From Eq. (\ref{93}), the second stage of the 
 evolution described here will proceed until:
\be
\xi=\xi_f={N^+-N^-\over 2N}+{\cal O}(\alpha)\ee
which implies:
\begin{eqnarray}
&&X=X_f={N^-\over N}+{\cal O}(\alpha)\\
&&\nonumber\\
&&T^{\pm}=T^{\pm}_f={2E_0\over Nk_B}+{\cal O}(\alpha)\\
&&\nonumber\\
&&\widetilde{\Delta}_2={k_BT^P_f\over 2m}
={E_0\over mN}+{\cal O}(\alpha)\hskip 5mm{\rm i.e.}
\hskip 3mm T^P_f=T^{\pm}_f+{\cal O}(\alpha)
\\
&&\nonumber\\
&&\widetilde{V}={\cal O}(\alpha)\\
&&\nonumber\\
&&\widetilde{\Pi}={\cal O}(\alpha)\label{114}
\end{eqnarray}
 At this point,  corrections in $\alpha^2$ have to be included;
the relevant perturbation approach
to get the further evolution  would use the rescaled
time $\tau_2\doteq \alpha^2t$ and consider also $\Delta_3$
and $\Delta_4$, and so on for the successive stages
of the relaxation towards complete equilibrium
(Maxwellian distributions for the gas particles).

\section{Remarks}

\noindent {\bf 1)} \  Our main result Eq. (\ref{main}) gives 
an equation totally independent of the parameters of the problem,
 except for the ratio $2N^{\pm}/N$, which will be 1 in our simulations.

\vskip 3mm
\noindent {\bf 2)} \ The expression (\ref{92}) for $V$
and the equation (\ref{93}) describing the evolution, 
coincide with the equations obtained in \cite{frachebourg}
using qualitative arguments. 

\vskip 3mm
\noindent {\bf 3)} \ The equation  (\ref{main}) for $\xi(s)$
was also derived  by J.L. Lebowitz \cite{joel}
using either qualitative arguments or the fact that the 
evolution of the piston
will be described by an Ornstein-Uhlenbeck process.

\vskip 3mm
\noindent {\bf 4) } \
The consistency between the short time solution
(obtained in \cite{GPL} with $\alpha=0$) and the solution
for large time, here obtained 
by considering evolution equations
with respect to rescaled variable $\tau$ 
and restricted at first order in $\alpha$,  
amounts to match the infinite-time limit
$t\rightarrow\infty$ of the zero-order solution
with the initial condition in $\tau=0$ of the
rescaled perturbation large-time solution,
which makes sense only for large $R$.

\vskip 3mm
\noindent {\bf 5) } \
From Eq. (\ref{92}), we see 
that $\widetilde{\Pi}>0$ if and only if $T^+>T^-$.
As we have seen above, it is precisely the condition
for $\widetilde{V}>0$.
Therefore work is delivered to the warmer side
but more heat is extracted from this warmer side so that
its energy decreases.

\vskip 3mm
\noindent {\bf 6)} \ 
We should compare the expression obtained for
the velocity Eq. (\ref{92}):
\be
V={2\over 3M}\;
\sqrt{8k_Bm\over \pi}\;
(\sqrt{T^+}-\sqrt{T^-})\ee
with the expression obtained in \cite{PG}
for the stationnary state of the piston in an infinite 
cylinder under the condition that the pressures on both sides
are equal:
\be
V_{st}={1\over M}\;
\sqrt{\pi k_Bm\over 8}\;(\sqrt{T^+}-\sqrt{T^-})\ee
The velocity $V$ is larger than
$V_{st}$ by a factor $16/3\pi$. It is 
related to the fact that in the present case, the pressures are not equal
but:
\be
p^--p^+={2m\widetilde{\Pi}\over M}+
{\cal O}(\alpha^2)\geq 0\ee

\section{Numerical simulations}

In order to check the assumptions 1, 2 and 3
and to test our main
results,  in particular
Eqs. (\ref{93})-(\ref{114}), we have conducted a large number of simulations.
We have considered a one-dimensional system of fixed
length $L$. 
The mass of the fluid particles is $m=1$ and the mass
$M$ of the piston is varied in the different simulations.
Initially the piston is at the position $X_0$ with
velocity $V_0=0$ and a configuration of $N^-$ and $N^+$
particles is taken at random
from Maxwellian distributions with 
temperatures $T_0^-$ and $T_0^+$. The system is then led to evolve 
according to the law of purely elastic collisions Eq. (\ref{coll}).
The position $X$ of the piston
is recorded, together with the temperatures and pressures in the
left and right compartments defined by the average (kinetic)
energy and the equation of state  $p^{\pm}=\rho^{\pm}k_BT^{\pm}$.

To conduct these numerical simulations, one is confronted with
the following dilemma. In order to reduce the fluctuations 
to a minimum to obtain significant results, one should take $N^{\pm}$
very large. This implies that $M$ should also be very large in order to have
a strong damping. However since the second stage proceeds 
on a time scale $\tau=2t/M$, this implies that one has to follow an 
enormous number of collisions, which needs a very long computer time.
For this reason, to study the first stage of the evolution, we have considered
up to 2 millions particles, but we were forced to reduce drastically this
number of particles to investigate the second stage, i.e. to approach thermal
equilibrium.
In all simulations presented below we have taken:
\vskip 1mm
$k_B=1, \hskip 7mm m=1, \hskip 7mm L=60, \hskip 7mm
N^-=N^+=\bar{N}=\displaystyle {N\over 2}$
\vskip 1mm
$X_0=10, \hskip 7mm V_0=0, \hskip 7mm T_0^-=1, \hskip 7mm
T_0^+=10$
\hskip 5mm
i.e.  \ $\rho_0^-=5\rho_0^+$ \hskip 3mm
and \hskip 3mm $p_0^-=\displaystyle {p_0^+\over 2}$
\vskip 2mm
\noindent
For these initial conditions, the end of the first stage,
i.e. mechanical equilibrium, is characterized by Eqs. (\ref{50})-(\ref{Xad})
which give:
\be\label{109}
X_{ad}=8.42, \hskip 7mm 
T_{ad}^-=1.54,\hskip 7mm T_{ad}^+=9.46,\hskip 7mm 
p_{ad}^-=p_{ad}^+=0.1833\;\bar{N}\ee
We concluded in \cite{GPL} that in the first stage,  the motion is
deterministic and depends strongly on $R=\bar{N}/M$
if $R<1$,  but become independent of $R$ if $R>4$.
In particular for $R<1$, it is weakly damped and the period of 
oscillations we have computed coincide with the period obtained 
from simulations and from hydrodynamics assuming 
adiabatic oscillations.
On the other hand for $R>10$, the motion is strongly
damped and is independent  of $N^{\pm}$ and $M$ as soon
as $M$ is large enough. Moreover
at the end of the first stage, after mechanical equilibrium has been reached,
simulations shows that oscillations
start to appear. These oscillations were interpreted as a consequence
of the fact that the velocity distributions in the fluid were not
Maxwellian.

In Figure 1, we have plotted the
first part of the evolution, i.e. the adiabatic
approach towards mechanical equilibrium, in function of time $t$ for 
$N^{\pm}=2.10^4$, $M=10^5$ (i.e. $R=0.2$, weak damping)
and for $N^{\pm}=3.10^5$, $M=3.10^4$ 
(i.e. $R=10$, strong  damping).
From the simulations the values obtained at mechanical equilibrium
are:
\be\label{110}
 X_{obs}=8.33\pm 0.05, \hskip 7mm 
T_{obs}^-=1.52\pm 0.04,\hskip 7mm T_{obs}^+=9.48\pm 0.04,\hskip 7mm 
p_{obs}^-=p_{ad}^+=0.183\;\bar{N}\ee
in good agreement with the analytical values Eq. (\ref{109}).
Other graphs for this adiabatic evolution can be found in \cite{GPL}.

In Figures 2-8, we have plotted the
second part of the evolution, i.e. the approach towards thermal equilibrium
in terms of the
 scaled variable $\tau=2t/M$,
obtained from our equations and from simulations  with 
$N^{\pm}=3. 10^4$ and $M=100, \; 200,\; 1000$.
The position in function of $\tau$ is presented in Figure~2, where no significant difference can be seen 
 for the different values of $M$. Figure~3
shows a zoom around $\tau=0$ and $\tau =16$
 for $M=200$ and $M=1000$, expressed in
terms of the real time $t=\tau M/2$. 
It shows that on the time interval
$(t,t+300)$ considered in this zoom, the evolution
(fast local relaxation slaved to the slow evolution of the system)
will tend to be
independent 
of $M$ for $M$ sufficiently large 
(fluctuations around mechanical equilibrium)
while on the large scale the evolution is
scaled by $1/M$, i.e. appears independent of $M$ when expressed
in function of $\tau=2t/M$.
In Figure~4,  we present the temperatures $T^{\pm}$ together
with the surface temperatures $T^{\pm}_{surf}$
computed from the incoming particles at distance 1 from the
piston. In Figure~5 the temperature of the piston is plotted using either
Eq. (\ref{101}) i.e. $T^P = \sqrt{T^+T^-}$
or the definition (\ref{76}) with $k_BT^P= M \Delta_2$; one should note
that using the definition (\ref{76}) implies a factor which
 is in our simulations
$30 000 (T^+  + T^-  -11)$ and thus the fluctuations are very large.
The other 
scaled quantities of order 1, i.e. 
$\widetilde{\Pi}=M(p^--p^+)/2$ and $\widetilde{V}=MV/2$ 
(recall that $m=1$) obtained from the simulations
and our Eqs. (\ref{92}) are represented in Figures 6 and 7.
Finally, since the distributions of velocities of the gases do not
remain Maxwellian, we compare in Figure 8 the previous evolution
with the evolution starting from the (adiabatic) initial conditions
(\ref{110}) with Maxwellian distributions. We observe no significant difference.





\section{Conclusions}

Using a two-time-scale perturbation approach, we have shown that
the evolution of a piston with large but finite mass proceeds in two stages.
For $R>10$, there is strong damping and in the first stage, at times
$t={\cal O}(1)$, the evolution is similar to the one previously
discussed in the thermodynamic limit \cite{GPL}:
it is a deterministic, adiabatic evolution
towards mechanical equilibrium.
At the end of this first stage,  adiabatic oscillations  appear
which are interpreted as associated with the fact that the
 velocity distributions of the gas particles are not Maxwellian.
In particular, simulations show that in the adiabatic evolution:
\be
{1\over T^-}\;{dT^-\over dt}\approx \;-\; 1.8\; {1\over X} \;{dX\over dt}\ee
In the second stage, at times $t={\cal O}(M)$,
a fluctuation-driven regime develops, leading to a
 relaxation towards thermal equilibrium.
In this regime, the motion of the piston is slaved
(i.e. it adapts on a time scale ${\cal O}(1)$)
to the slowly relaxing (on a time scale ${\cal O}(M)$)
asymmetry of the thermal fluctuations
on each side.
Moreover, since the motion of the piston is now stochastic, one has to 
introduce the temperature of the piston. Therefore the fluctuations in the 
piston motion produce a heat transfer from the warmer to the colder
side which is larger than the work produced by the piston motion.
In particular, in this stage, simulations as well as equations show that now:
\be
{1\over T^-}\;{dT^-\over dt}\approx\;\;   {1\over X}\; {dX\over dt}\ee
In this regime, the piston is no longer adiabatic but the time 
involved to reach thermal equilibrium is of order $M$.

\vskip 3mm
In the first stage, simulations are in very good
agreement with the predicted values for the position of
 the piston and the temperatures of the gases at mechanical equilibrium.
Moreover for weak damping ($R<1$) the observed values for the period
coincide within a few per cent with the predicted values from our
equations, as well as with the values obtained from thermodynamics
assuming ``adiabatic oscillations''.
However the observed values
 for the damping coefficient disagree strongly with 
the computed values. Unravelling the origin of the friction,
we must conclude that there are two different friction coefficients. 
A first coefficient describes the stabilization of the piston velocity
in the infinite length situation, or for the case of finite length
(and $R>10$) until the first recollision on the piston takes place.
It is related to the shock waves propagating uniformly on both sides of the surface
of the piston. The second coefficient describing the damping of 
the oscillations is associated with the rebouncing of these shock 
waves on the surface of the cylinder and of the piston. It appears 
to be related to the relative velocity of the shock waves with respect
 to the piston velocity \cite{morris}.

For the second stage, as can be seen on Figures 2 to 7, our main
conclusions are all verified by numerical simulations
 with high degree of accuracy.
One should however observe that the agreement would
be ``perfect'' if we would consider, instead of Eq. (\ref{main}),
the equation:
\be
{d\xi\over ds}=-\;0.8\;\left[
\sqrt{{N\over 2N^+}(1+2\xi)}\;
-\;\sqrt{{N\over 2N^-}(1-2\xi)}\;
\right]\ee
The origin of the discrepancy could be traced to our  Assumption 3b
where we use the Maxwellian values for $F_1^{\pm}(V=0)$.
It is known \cite{frachebourg} that at this stage the distribution functions
$f^{\pm}(v)$ differ considerably from
the Maxwellian distributions around $v=0$.
Of course this discrepancy could also come from
the average assumption (Assumption 2);
however from Figure 4, we see that the surface temperatures oscillate
around the bulk temperatures (on a time scale $t$) and thus
 we expect this technical
assumption 
to have  no important consequences.

\vskip 3mm
Let us remark that if we compute the 
time auto-correlation function $C(t)$
of the position $X$ of the piston and  the associated
power spectrum $S(\omega)$ (the Fourier transform of $C(t)$), then from
the simulation we could 
  expect a three-peak structure as 
observed in the equilibrium state by  White et al. \cite{white}
recovering a standard result  in hydrodynamics \cite{hansen}.
The central peak describes 
(according to the fluctuation-dissipation theorem) both

-- the (slow) relaxation towards thermal equilibrium,
described by a term $\sim e^{-t/\tau_{th}}$
in the time auto-correlation function $C(t)$, with  $\tau_{th}\sim 
1/\alpha \sim M$; the width of the peak is ${\cal O}(1/M)$.

-- and the thermal fluctuations of the piston
as described by $\Delta_2(\infty)$.

\vskip 1mm
\noindent
The two lateral peaks, slowly evolving in time
with $X(\tau)$ and $T^{\pm}(\tau)$
correspond to the fast adiabatic and deterministic
relaxation towards the  instantaneous mechanical equilibrium,
slaved to the values of  $X(\tau)$ and $T^{\pm}(\tau)$.
It also gives (fluctuation-dissipation theorem in this
quasi-equilibrium state) a contribution to the fluctuations
of the piston motion. The width of these peaks
is independent of $M$.

\vskip 3mm
The perturbation approach can be carried on at higher orders;
it would give access to the following stages of the relaxation,
at increasing time scales ${\cal O}(1/\alpha^n)\sim {\cal O}(M^n)$,
involving higher moments $\Delta_{2n-1}$
and $\Delta_{2n}$.
Maxwellian distributions for the velocity distributions inside the
gases would be reached 
only in the final stage s of this relaxation (namely at times
$t\sim {\cal O}(M^n)$ with $n$ large ). 
However the factorization property will be violated at these orders and we can not say
what would be the consequences of this violation on the results
obtained within our Boltzmann's equation approach.

\vskip 20mm
\noindent
{\bf Acknowledgements:}
We are grateful Prof. E. Lieb for many discussions on the second law for
adiabatic systems. We are also grateful
 to Prof. J.L. Lebowitz
for sending us his results before publication, and
 to Prof. N. Chernov for a long series of e-mail discussions.
A. Lesne greatly acknowledges the hospitality at the
Institute of Theoretical Physics of the \'Ecole Polytechnique F\'ed\'erale
de Lausanne where part of this research has been performed.
Finally, we thanks the ``Fonds National Suisse de la Recherche Scientifique''
for his financial support of this project.


\newpage
\noindent
{\Large\bf Figure captions}

\vskip 15mm
\noindent
{\bf Figure 1:}  Adiabatic stage of the evolution for 
$T_0^-=1$ and $T_0^+=10$.

a) strong damping: $N^{\pm}=3.10^5$, $M=3.10^4$, $R=10$.

b) weak damping: $N^{\pm}=2.10^4$, $M=10^5$, $R=0.2$.

\vskip 5mm
\noindent
{\bf Figure 2:}  Approach to thermal equilibrium for 
$N^{\pm}=3.10^4$ and $M=100, 200, 1000$ compared with the solution
of Eq. (\ref{93})  (the real time is $t=\tau M/2$).

\vskip 5mm
\noindent
{\bf Figure 3:}  Zoom on the evolution of Figure 2 in function 
of the real time $t$ for $M=200$ (light curve) and $M=1000$
(bold  curve) around $\tau=0$ and $\tau=16$.

\vskip 5mm
\noindent
{\bf Figure 4:} 

a) Evolution of the temperatures $T^{\pm}(\tau)$
for $N^{\pm}=3.10^4$ and $M=200$ compared with Eqs. (\ref{93}) and (\ref{100});

b) Surface temperatures $T^{\pm}_{surf}(\tau)$
compared with bulk temperatures $T^{\pm}(\tau)$.

\vskip 5mm
\noindent
{\bf Figure 5:} Temperature of the piston
($N^{\pm}= 3.10^4$, $M=200$)

a) from $T^P =\sqrt{T^+T^-}$
(Eq. (\ref{101}) with $T^{\pm}$ obtained from Eq. (\ref{93}) (light
curve);

 b) from $T^P =\sqrt{T^+T^-}$ with $T^{\pm}$ obtained from the
simulations (dark curve);

  c) from the definition Eq. (\ref{76}) with $k_BT^P= M\Delta_2$ (stochastic curve).

\vskip 5mm
\noindent
{\bf Figure 6:} Scaled pressure difference
$A\widetilde{\pi}=A(p^--p^+)M/2$ 
computed
from Eq. (\ref{101})   with $T^{\pm}$ obtained from Eq. (\ref{93}) and from
simulations  (with $N^{\pm}= 3.10^4$, $M=200$).

 .

\vskip 5mm
\noindent
{\bf Figure 7:} Scaled velocities ($N^{\pm}= 3.10^4$, $M=200$)

a) from Eq. (\ref{93}) (light curve);

b) from Eq. (\ref{101})  with $T^{\pm}$ obtained from the simulations
(dark curves);

c) average velocity obtained from the simulations.

\vskip 5mm
\noindent
{\bf Figure 8:} Evolution for $N^{\pm}=3.10^4$, $M=200$
compared with the evolution starting from initial conditions Eq. (\ref{110})
and Maxwellian velocity distributions.

\newpage
\hspace*{20mm}
\leavevmode
\epsfxsize= 90pt
\epsffile[10 15 200  700]{adpos.eps}

{\bf Figure 1}
\newpage
\hspace*{20mm}
\leavevmode
\epsfxsize= 90pt
\epsffile[10 15 200  700]{pos.eps}

{\bf Figure 2}
\newpage
\hspace*{50mm}
\leavevmode
\epsfxsize= 40pt
\epsffile[100 150 200  700]{zoom0.eps}
\hspace*{15mm}(a)
\vspace*{-10mm}

\hspace*{50mm}
\epsfxsize= 40pt
\epsffile[100 150 200  700]{zoom16.eps}
\hspace*{15mm}(b)
\vspace*{-20mm}
{\bf Figure 3}

\newpage
\hspace*{50mm}
\leavevmode
\epsfxsize= 40pt
\epsffile[100 150 200  700]{Temp.eps}
\hspace*{15mm}(a)
\vspace*{-10mm}

\hspace*{50mm}
\epsfxsize= 40pt
\epsffile[100 150 200  700]{Tempsurf.eps}
\hspace*{15mm}(b)
\vspace*{-5mm}

{\bf Figure 4}

\newpage
\hspace*{20mm}
\leavevmode
\epsfxsize= 90pt
\epsffile[10 15 200  700]{Tp.eps}

{\bf Figure 5}\newpage
\hspace*{20mm}
\leavevmode
\epsfxsize= 90pt
\epsffile[10 15 200  700]{Api.eps}

{\bf Figure 6}\newpage\hspace*{20mm}
\leavevmode
\epsfxsize= 90pt
\epsffile[10 15 200  700]{Vp.eps}

{\bf Figure 7}\newpage
\newpage\hspace*{20mm}
\leavevmode
\epsfxsize= 90pt
\epsffile[10 15 200  700]{comparaison.eps}

{\bf Figure 8}\newpage

 \end{document}